# Ion-Sound Waves in UHF Discharge Plasma Injected in Open Magnetic Trap

S. Nanobashvili[1], Z. Beria[1], G. Gelashvili[1], G. Gogiashvili[1], I. Nanobashvili[1],

G. Tavkhelidze[1], I. Tsevelidze[1], G. Van Oost[2]

1. Andronikashvili Institute of Physics of the Iv.Javakhishvili Tbilisi State University, Tbilisi, Georgia
2. Department of Applied Physics, Ghent University, Ghent, Belgium

1. Introduction

We investigate in stationary regime on the linear plasma device −OMT-2 [2,3] the methods of plasma heating by electromagnetic waves, interaction of electromagnetic waves with magnetoactive plasma and plasma turbulence and transport processes in the trap. We use ultra-high frequency (UHF) contactless method to fill the open magnetic trap by plasma. We have proposed new method of open magnetic trap filling with plasma −plasma is injected in the trap along the magnetic field from independent stationary UHF source where plasma is formed in ECR regime. In this paper we present the results of investigation of magnetic trap filling by plasma and the results of investigation of plasma low frequency (LF) oscillation in the trap.

2. Experimental set-up

Experiments were carried out on a stationary installation, the diagram of which is presented in Fig.1. It consists of two main parts: an independent UHF plasma source and open magnetic trap, in which plasma is injected.

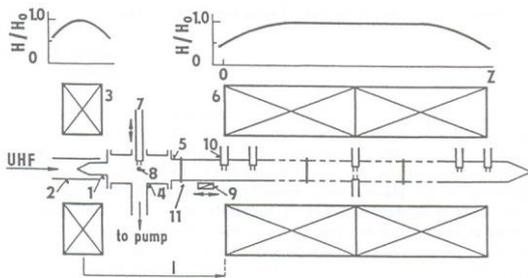

Fig.1 The scheme of experimental setup.

1 - discharge chamber, 2 −rectangular waveguide,
3 −coil forming the magnetic field in source,
4 −diagnostic section, 5 −volume under investigation, 6 - solenoid, 7,10 − double electric probes, 8,9 − semiconducting light sensors. 11 −capacitance probes

In the UHF source plasma is formed in a quartz tube 1 with inner diameter 2.6 cm and length of 10 cm. Stationary UHF power ( 2400 MHz, 150 W) is supplied to the discharge chamber by standard rectangular waveguide 2. The discharge chamber with the waveguide is located into the stationary magnetic field, created by a short coil 3. Magnetic field in the center of the coil can be continuously changed from zero to the maximum value of 2000 Oe.

The discharge chamber of the UHF plasma source is connected with the cylindrical section 4, made of stainless steel and being a diagnostic section [3].

The investigation volume 5 with plasma is placed into the stationary magnetic field, created by solenoid 6 with inner diameter of 19.5 cm and the length of 90 cm. By solenoid we obtain the uniform magnetic field, as well as the field of mirror and multimirror configuration with controllable mirror ratio. The maximum field along the axis of the solenoid can be varied smoothly from zero to maximum value of 5000 Oe. The results presented in the this paper deal with the plasma injection into the trap with uniform magnetic field. Injected plasma parameters - density of charged particles, temperature of electrons and their distribution over the radius - were measured by movable double electric probe 7.

In order to determine the efficiency of filling the magnetic trap with plasma semiconducting light sensor 9 moving along the chamber 5 were used. In our case plasma is weakly ionized. As it is well known integral emission of plasma depends on its particle density. We have verified this experimentally. For determination of the spectrum of plasma oscillations we used capacitance probe 11.

During experiments described above Argon and Helium were used as working gas. Plasma in the UHF source was created at working gas pressure in the range $10^{-5}$ - $10^{-2}$ Torr.

3. Results and discussion

1) <u>UHF plasma source</u> - discharge in the source can be obtained in the investigated range of working gas pressure by means of stationary UHF power only if the condition of electron cyclotron resonance is fulfilled in the region of UHF field interaction with plasma, $\omega_0 = \omega_{He} = eH/mc$, i.e. when magnetic field equals to that of cyclotron ($H_c = 850$ Oe). The measurements by double electric probe 7 have shown that under our experimental conditions plasma density can be changed from $10^8$ to $10^{12}$ cm$^{-3}$ with plasma electron temperature $T_e = 2 - 3$ eV. These changes are quite controllable and well reproducible.

2) <u>Plasma injection in magnetic trap</u> - plasma with above given parameters has been injected from UHF source into uniform magnetic field and magnetic trap of mirror configuration formed by a solenoid 6 (Fig.1). Filling of open magnetic trap with plasma has been studied experimentally in wide range of parameters of the injected plasma, neutral gas prsesure, magnetic field strength in the trap and also the distance betvin plasma and solenoid.

Tipical results of experimental determination of the efficiency of open magnetic trap filling by plasma

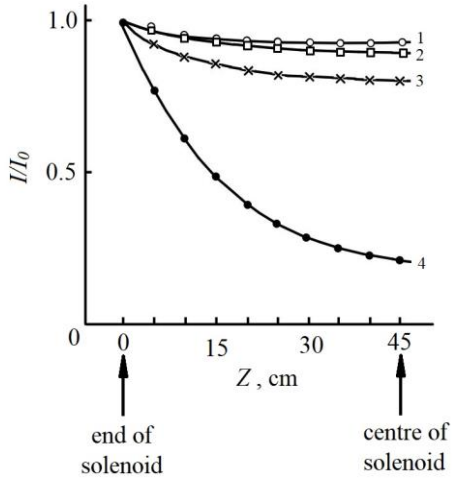

Fig.2
1 - p=2.3·10⁻⁴ Torr;  2 - p=4.6·10⁻⁴ Torr;  3 — p=7.6·10⁻⁴ Torr;  4 - p=2.3·10⁻³ Torr;
for  $H_t$=400 Oe

under different pressure and magnetic field are presented on Fig.2 and Fig.3.

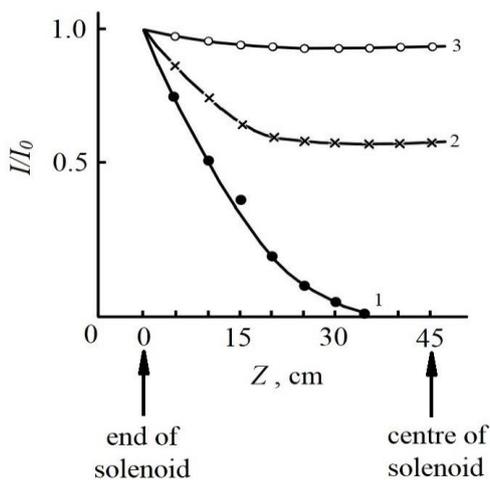

Fig.3
1 - $H_t$=0; 2 - $H_t$=100 Oe; 3 - $H_t$=400 Oe;
for  p=2.3·10⁻⁴ Torr

Our experiments showed that for pressure p < 5·10⁻³ Torr, magnetic field in the trap $H_t$ < 400 Oe and distance between the plasma source and solenoid $\ell$ < 80 cm the filling of the trap is very effective and rather quiescent plasma with controllable density within the range $10^8$ - $10^{12}$ cm⁻³ and temperature 2 - 3 eV is accumulated in the trap.

3) **Low frequency oscillations** of plasma in magnetic trap - under the above mentioned experimental conditions we have found that near the UHF plasma source and also in the trap first and second harmonics of ion sound wave are reliably detected.

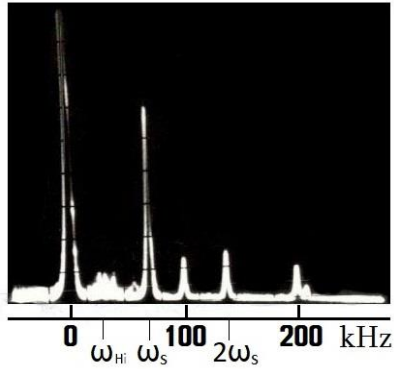

Fig.4 Plasma oscillation spectrum in in Ar

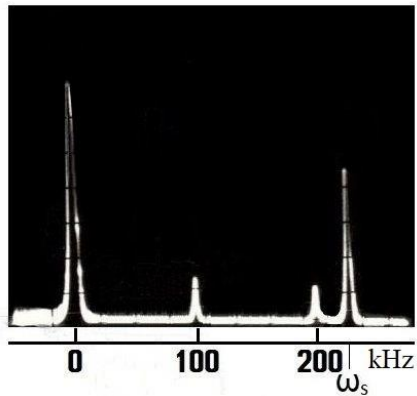

Fig.5 Plasma oscillation spectrum in in He

It is known that in non-equilibrium magnetoactive plasmas where the condition $T_e \gg T_i$ is fulfilled ion acoustic oscillations with frequency

$$\omega_s = \frac{\pi\, l}{L} \cdot \left(\frac{3kT_e}{M_i}\right)^{1/2} \qquad (1)$$

($l$ is oscillation mode number, $L$ - characteristic size of plasma and $T_e$ - electron temperature of plasma) may appear.

Estimations show that oscillations observed in Argon at a frequency of the order of ~70 kHz, cal-culations according to the formula (1) for our experimental conditions ($T_e$ = 2 - 3 eV, $L$ = diameter of discharge chamber and $l = 1$) correspond well to the value of experimentally measured ion sound waves with wavelength equal to the inner diameter of the discharge chamber.

Indeed, in the same conditions as in Argon in Helium plasma we detect oscillations with frequency ~210 kHz, which corresponds exactly to ion sound oscillation frequency (1) shifted by a factor of square root from Argon and Helium mass ratio.

This is an additional argument in support of the statement that under our eqsperimental conditions excitation of ion sound oscillations takes place in plasma.

4. Conclusion

The presented experimental results allow to conclude that the method of magnetic trap filling with plasma proposed by us allows to simply accumulate target plasma in open magnetic trap where first and second harmonics of ion sound wave are reliably detected.